# The Pace of Artificial Intelligence Innovations: Speed, Talent, and Trial-and-Error[1]

Xuli Tang
School of Information Management, Wuhan university, Wuhan, Hubei, China

Xin Li (lucian@whu.edu.cn)
Information retrieval and Knowledge Mining Laboratory, School of Information Management, Wuhan University, Wuhan, Hubei, China

Ying Ding
School of Information, The University of Texas, Austin, Texas, U.S.A.
Dell Medical School, The University of Texas, Austin, Texas, U.S.A.

Min Song
Department of Library and Information Science, Yonsei University, Seoul, Korea

Yi Bu
School of Informatics, Computing, and Engineering, Indiana University, Bloomington, IN, U.S.A.

*: Correspondence concerning this article should be addressed to Xin Li. The address of corresponding author is School of Information Management, Wuhan University, Wuhan430072, Wuhan, Hubei, China.

**Abstract**

Innovations in artificial intelligence (AI) are occurring at speeds faster than ever witnessed before. However, few studies have managed to measure or depict this increasing velocity of innovations in the field of AI. In this paper, we combine data on AI from arXiv and Semantic Scholar to explore the pace of AI innovations from three perspectives: AI publications, AI players, and AI updates (trial and error). A research framework and three novel indicators, Average Time Interval (ATI), Innovation Speed (IS) and Update Speed (US), are proposed to measure the pace of innovations in the field of AI. The results show that: (1) in 2019, more than 3 AI preprints were submitted to arXiv per hour, over 148 times faster than in 1994. Furthermore, there was one deep learning–related preprint submitted to arXiv every 0.87 hours in 2019, over 1,064 times faster than in 1994. (2) For AI players, 5.26 new researchers entered into the field of AI each hour in 2019, more than 175 times faster than in the 1990s. (3) As for AI updates (trial and error), one updated AI preprint was submitted to arXiv every 41 days, with around 33% of AI preprints having been updated at least twice in 2019. In addition, as reported in 2019, it took, on average, only around 0.2 year for AI preprints to receive their first citations, which is 5 times faster than 2000–2007. This swift pace in AI illustrates the increase in popularity of AI innovation. The systematic and fine-grained analysis of the AI field enabled to portray the pace of AI innovation and demonstrated that the proposed approach can be adopted to understand other fast-growing fields such as cancer research and nano science.

## 1. Introduction

It is irrefutable that innovations in artificial intelligence (AI) are occurring at a rate faster than ever before. The chess-playing program, as an important testbed for AI, was originally written in 1951 (Bard et al., 2020), but it took almost 50 years to develop the DeepBlue system into one capable of beating the world champion Garry (Campbell, Hoane, & Hsu, 2002). However, in light of big data and high-performance computing advancements,

AI in chess has now reached superhuman performance, all within the short timeframe of 10 years (Wang et al., 2016). In a similar pattern, the studies on artificial neural network (ANN), once stagnant for nearly 60 years after the original Perceptron was introduced in 1957 (LeCun, Bengio, & Hinton, 2015), have now, in the last 10 years, prompted AI into a new era through deep learning (DL) and been applied to a variety of industries, from unmanned aerial vehicles (Kattenborn, Eichel, & Fassnacht, 2019), to precision medicine (Oakden-Rayner et al., 2017), and smart finance (Heaton, Polson, & Witte, 2017).

Entering the 21st century, espeically in light of recent years, everyone can feel the gravitational pull of AI. Newly rejuvenated, AI itself is an efficient, effective, and intelligent technological tool—one that is inexorable in driving humanity into a new era. Everyday, science fiction scenarios of advanced civilizations merge further and further with reality. The speed at which AI breakthroughs have appeared can be evidenced in both the academic and industrial field. They are a contributor to the acceleration of many technological breakthroughs, from recurrent neural nertworks for natural language understanding (Varghese et al., 2018), to computer visions for medical imaging diagnosis (Doan & Carpenter, 2019), and deep learning for antibiotic discovery (Strokes et al., 2020). Investigating the pace of AI innovations has thus become increasingly relevant. Yet, few studies have managed to measure or depict this increasing velocity of innovations in AI.

Innovation speed, originally proposed by Kessler and Chakrabarti (1996), has been widely used as a concept for monitoring product innovation, varying in different contexts (Cankurtaran, Langerak, & Griffin, 2013; Carbonell & Rodriguez, 2006; J. Chen, Damanpour, & Reilly, 2010; Kessler & Chakrabarti, 1996). The time interval for a product to evolve from an initial idea to a marketable entity (Kessler & Chakrabarti, 1996), frequency of product updates (Dong, Wu, & Zhang, 2019) , and diffusion rate of a product in an organization (Rogers, 2010) can all be measured through innovation speed. Citation counts and publication rate can be calculated as well. Take, for example, Dushnitsky and Lenox (2005), who adopted a firm's citation-weighted count of patents as a fitting surrogate for the rate of innovation within firms. Agarwal and Searls (2009) captured occurrences of innovations as well, using the publication record for the end product, eventually proving that novel therapies lagged behind the initial scientific discoveries by years.

Rapid innovations are commonly defined as exploring possible opportunities in a limited time-interval (Banu Goktan & Miles, 2011), or trial and error with an abundant timeframe (Kessler & Chakrabarti, 1996).

Similar to the mantra "fail fast and fail often," trial and error plays a crucial role in the success of innovations that tend to have shorter and more frequent trial-and-error lifecycles. It has become an important feature in measuring the pace of innovation. Based on previous studies on arXiv (Brody, Harnad, & Carr, 2006; Larivière et al., 2014) , we can measure the trial and error by the process of refining, improving and updating the current version of AI preprints.

Publications in both conferences and journals are one of the most common forms of innovation in AI studies (Frank et al., 2019). Open academic datasets, such as MAG and Semantic Scholar, include massive publications on AI with bibliographic information, which can serve as reliable data sources for identifying the pace of AI innovations (Deng, 2018; Frank et al., 2019). Meanwhile, as the free and rapid dissemination mechanizes, preprints in arXiv form another important composition of AI innovation. Specifically, the fine-grained metadata indexed in arXiv, such as submission time and version update time, enable us to study AI innovation pace up to the seconds (Larivière et al., 2014). Hence, with these datasets, we are capable of capturing the pace of AI innovations through the following perspectives: the time intervals of newly emerged AI innovations, the fast growth of AI research working force, and the rate of trial and error in AI innovations.

Understanding the pace of AI from the perspective of AI publications, AI players, and AI updates (trial and error) is the focus of this study. According to our analyses, in all these three aspects, the field of AI has entered a stage of rapid innovation. First, for AI publications, in each hour of 2019, 3.817 AI preprints were submitted to arXiv, over 148 times faster than in 1994. Furthermore, a deep learning–related preprint was submitted to arXiv every 0.87 hours in 2019, over 1,064 times faster compared than in 1994. Moreover, at least one AI preprint was submitted to arXiv per day from 2014–2019, while only 98 days (27% of a year) between 1993 and 1999 had any AI preprint submitted.

Second, in terms of AI players, 5.26 new researchers entered into the field of AI every single hour in 2019, about 175 times faster than in the 1990s. There were 46,097 new researchers that entered into the field of AI in arXiv in 2019 alone.

Third, in terms of AI updates, an updated version of the AI preprint was submitted to arXiv every 41 days, with around 33% of arXiv's AI preprints being updated twice or more in 2019. In addition, it took about 0.2 year on average for AI preprints to get their first citations from 2014–2019, around 5 times faster than from 2000–2007.

## 2. Related work

### *2.1. Innovation speed*

Innovation speed is one of the detectors in predicting the leadership of a country, university, or company, especially for new discoveries in the field of AI. Most of the extant studies on innovation speed are conducted in the field of new production development (NPD), in which innovation speed is defined as the time to market of the first product. These studies can be summarized into three categories: the product development perspective, the investment perspective, and the communication perspective.

Research from the product development perspective provides insight into how firms speed up their innovation activities by exploring the association between innovation speed and product characteristics, such as product quality, performance, efficiency, and success. Take for example, Kessler and Bierly (2002), who analyzed the implications of innovation speed in 75 product development projects and concluded that there was a positive association between faster innovation speed and higher product quality. Allocca and Kessler (2006) demonstrated that the innovation speed had a synergistic effect on the projects' quality, efficiency, and success. Carbonell and Rodriguez (2010) conducted a questionnaire survey of employees in the manufacturing industry and found that the increase in a company's innovation speed improves its new product performance.

The investment perspective emphasizes the importance of resource allocation and portfolio management for firms to accelerate their innovation speed. Proper R&D structure and strong human resources are considered as the important internal resources affecting the innovation speed of a firm (Barney, 1991; Cainelli et al, 2015; Wernerfelt, 1984). Meanwhile, scientists also pointed out that external resources, such as collaboration with big firms, inter-organizational relationships, and networks, are significant for firms to enhance their innovative capacity and speed as well (Mikhailitchenko, & Lundstrom, 2006; Ireland et al., 2002). Additionally, hybrid resources, such as knowledge embodied in R&D services and patents from other firms, can contribute to the innovation process by offering firms useful external knowledge (Cainelli et al, 2015).

As for the communication perspective, the studies have focused on the association between innovation speed and team-related factors. For example, Allocca and Kessler (2006) proposed a conceptual framework of innovation speed for small and middle-sized enterprises, proceeding to validate it with 158 industrial projects. Heirman and Clarysse (2007) focused on start-ups and concluded that founder's experience and team tenure lead to faster innovation speed, while the collaboration with other companies had no significant effect on innovation speed. Except for studies from the aforementioned three perspectives, in the field of science of science, Zhai et al. (2018) depicted the patterns for the diffusion of innovation within the field of Latent Dirichlet Allocation (LDA) using a citation analysis that revealed how innovations are improved over time. Dong et al. (2019) treated the innovation speed of Open Source Software (OSS) as the updating speed and found that faster OSS innovation speeds could increase its downloading times.

In summary, previous studies have been mainly concerned about providing interpretations to speed up the innovation process of new production development in organizations. However, these studies have offered few insights into the speed of AI innovation.

*2.2. Factors affecting innovation speed*

Innovation speed is related to plenty of objective factors such as human resources and leaders' experience. In this paper, we use AI research as the proxy. Publications are one of the most common forms of innovation in AI research (Ahmadpoor & Jones, 2017; Fleming, Greene, Li, Marx, & Yao, 2019; Glorot, Bordes, & Bengio, 2011; Li, Azoulay, & Sampat, 2017). Therefore, we define the innovation speed in the AI contest as how soon a new AI publication (preprint) is released. The factors affecting innovation speed in AI research have been classified into two categories in this review: human capital and trial and error.

Innovation tends to depend heavily on human capital (Chen & Huang, 2009; Youndt et al., 1996). In the economy, the growth of a population can efficiently scale up innovation, even if individual contributions are found to be in decline (Jones, 1995). Similarly, the increase of new coauthors in a discipline can significantly boost innovation speed. Applying network approaches on coauthor teams, Wang and Hicks (2015) found that new coauthors will improve the innovation ability of the old team. Newcomers are proven to have improved innovation speed by introducing new knowledge, breaking the stereotype of thinking in repeated collaborative experience, and leading to diversified research perspectives (Gupta, Smith, & Shalley, 2006; Perretti & Negro, 2006). To better understand human capital and innovation, a more nuanced approach is needed.

Tolerance for failure spurs innovation. Compared to the National Institutes of Health (NIH), the Howard Hughes Medical Institute (HHMI) shows more tolerance for early failure, rewards long-term success, and gives its appointees great freedom to experiment. Azoulay, Graff, and Manso (2011) studied the careers of scientists funded by HHMI with a control group of similarly accomplished NIH-funded scientists and found that the HHMI-funded researchers produce high-impact articles at a much higher rate. Tian and Wang (2011) discovered that IPO firms backed by more failure-tolerant venture capital investors are significantly more innovative. It is unacceptable to omit trial and error when adopting measures as a proxy for innovation (Donoso, 2017). Trial and error can substantially better the innovation speed. The uncertainty and high risk are inherent qualities of innovation, leading to rich dynamics in creativity (Callander & Matouschek, 2019). The fast speed in error-trying can provide creators with more opportunities for exploration and improve the efficacy of the innovation process (Banu Goktan & Miles, 2011).

## 3. Methodology

In this paper, we define the pace of innovation in a research field is the speed at which the scientific innovation is occurring in the research field (Kessler & Chakrabarti, 1996; Rogers, 2010; Zhai et al., 2018). The overall process of the methodology is illustrated in Fig. 1. To portray the pace of AI innovations, the methodology is designed to be executed in four discrete steps. First, the data are collected from arXiv and semantic Scholar, the fields of which include arXivID, title, abstract, authors, submission information, topic words and update information, etc. The dataset is then stored in a local MySQL database. Second, AI papers and AI authors are classified into subgroups based on different classification schemes, to provide the nuanced details about the pace of AI innovations. Third, we propose three novel indicators, i.e., Average Time Interval (ATI), Innovation Speed (IS) and Update Speed (US), to quantify the pace of innovations. Finally, the pace of innovations in AI is analyzed from three perspectives: AI publications, AI players and AI updates (trial and error). The details on each step is as follows.

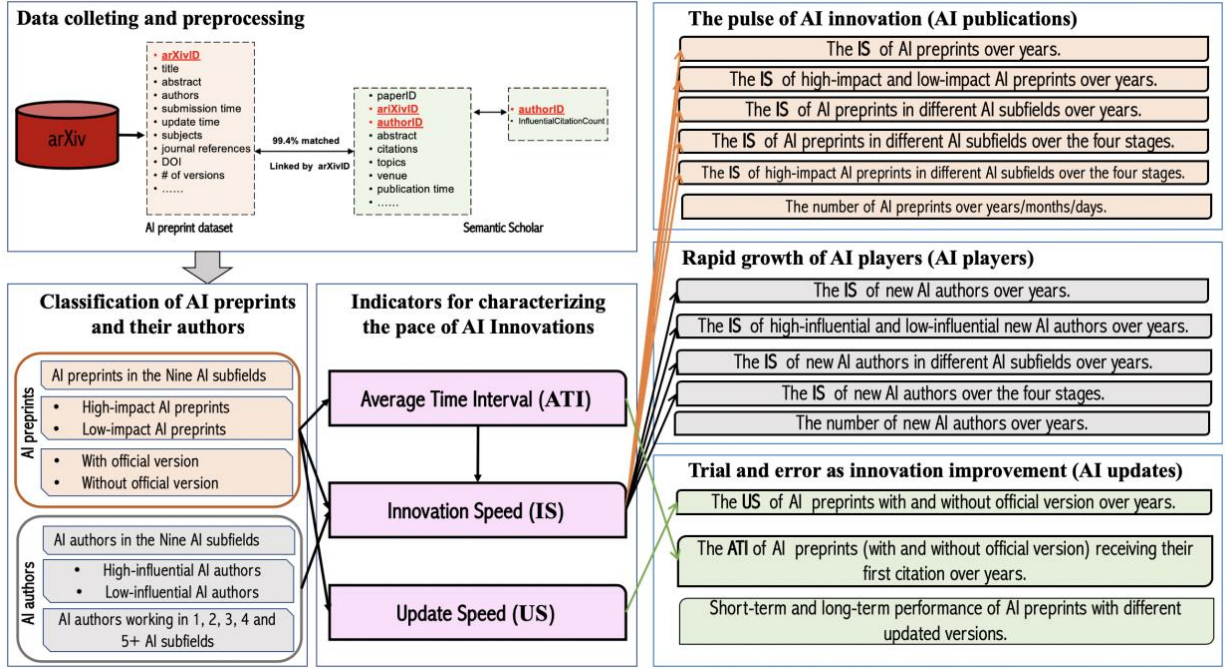

**Fig. 1.** The framework of this study.

*3.1. Data collecting and preprocessing*

As shown in the top left of Fig. 1, the data used in this study were collected from two sources: arXiv (https://arxiv.org/) and Semantic Scholar (https://www.semanticscholar.org/). The obtained data were pre-processed and reorganized into a dataset (i.e., AI preprint dataset) before being stored in a local MySQL database. The specific information on the data collection and preprocessing are provided as follows.

The AI preprint dataset is collected from arXiv, which is one of the largest and best-curated open-access archives for computer science. We used preprints submitted in the categories of AI, computation and language, computer vision and pattern recognition, machine reading, neural and evolutionary computing, robotics, multiagent systems, and information retrieval to represent AI research between 1993 and 2019 (Frank et al., 2019). In particular, a web crawler was developed with Python to download the AI preprints in HTML files from the arXiv website on February 6th, 2020. Then, we developed an HTML parser using the dom4j library in Java to extract the bibliographic information for each preprint, including arXiv ID, title, abstract, author names, submission history, subjects, DOI, and journal references. After removing the duplications, we obtained 117,509 unique records, which were stored in a local MySQL database for further analysis.

Next, we obtained the updated information from the submission history of preprints, such as the number of versions and the time for each submission and update, which were recorded at the level of seconds in arXiv and can be used to quantify the speed of AI innovations. Further, we linked the preprints to Semantic Scholar via their arXivID, to complete their bibliographic information, including author affiliations, author influence ("InfluentialCitationCount"), total citation counts, formal publication time, the time to get their first citations, and research topic words, etc. Six-hundred-and-ninety-six (0.06%) preprints that could not be found in the Semantic Scholar database were deleted. We also used the author identifier in Semantic Scholar (Fricke, 2018) to assign each author a unique ID for the remaining 116,813 preprints. The results indicated that 157,856 unique researchers authored these preprints.

It is also important to investigate the pace of AI innovations over the different development stages of AI research. Therefore, we divided the AI research in arXiv into four stages (He et al., 2019;Frank et al., 2019; Nilsson,2014; Russell & Norvig, 2016; Tran et al., 2019): the embryo stage (1993–1999), stable stage (2000–2007), machine-learning stage (2008–2003), and deep-learning stage (2014–2019).

### *3.2. Classifying AI preprints and their authors*

To comprehensively investigate the pace of AI innovation, we further classify AI preprints and their authors into groups.

(1) Classifying AI preprints

It is significant to compare the pace of innovations in different AI subareas. In this paper, we develop a new classification schema comprised of nine subcategories based on empirical work in content analysis and the previous studies (Annapureddy et al., 2020; Salatino et al., 2018), instead of directly using the subjects of preprints indexed by arXiv. We have two reasons for that. First, it is the 1998 ACM Computing Classification Scheme that is used for the classification of preprints in arXiv (https://arxiv.org/corr/subjectclasses). This old classification system can guarantee a relatively stable scheme that covers all of the computer science in arXiv; however, it can't group the AI preprints well since AI is a rather active and dynamic research area. Several important and active subareas of AI, such as deep learning and knowledge representation, have not yet been listed as subjects in the system. Second, some AI preprints could have been classified as "Other" when the authors could not find an appropriate subject area for it in arXiv.

Specifically, we first capture all research topic words of all preprints from Semantic Scholar. To simplify our analysis, these topic words were then assigned to nine subcategories: 1) natural language processing; 2) knowledge representation and reasoning; 3) planning and scheduling; 4) information retrieval; 5) robotics; 6) intelligent agents; 7) computer vision; 8) deep learning; and 9) machine learning. The detailed topic words of each category can be seen in Appendix A, and they were used as clue words to classify the AI preprints. We listed "deep learning" as an independent category rather than including it in the category of "machine learning," as we aim to highlight the recent deep-learning efforts in AI innovation. The distribution of preprints in different AI subfields is shown in Table 1. Note that one preprint can be classified into multiple subfields. For example, the preprint entitled "Image super-resolution using deep convolutional networks" (Dong et al. 2015) were classified into both the subcategories "deep learning" and "computer vision" according to its research topics. Multi-label classification makes it possible for us to analyze the pace of AI innovations in a systematic and fine-granular levels.

Moreover, to further investigate the innovation speed of preprints with different citation impacts, we categorized the AI preprints into high-impact and low-impact AI preprints based on their citation counts. Specifically, all the AI preprints were ranked in a descending order by citation counts, and the top 20% of them were considered as high-impact, last 40% as low-impact (Gayen, Bhavsar & Chandra, 2017). Besides, according to whether a preprint has been officially published in a journal or a conference, we also classify the AI preprints into two categories, i.e., AI preprints with official version and AI preprints without official version.

**Table 1**
The distribution of preprints in different AI subfields.

| No | Paper categories | Abbr. | # in AI preprint | Percentage (%) |
|---|---|---|---|---|
| 1 | Natural language processing | NLP | 16,126 | 13.8 |
| 2 | Knowledge representation and reasoning | KR | 6,319 | 5.4 |
| 3 | Planning and scheduling | PS | 5,086 | 4.4 |
| 4 | Information retrieval | IR | 2,514 | 2.2 |
| 5 | Robotics | RO | 1,540 | 1.3 |
| 6 | Intelligent agents | IA | 2,571 | 2.2 |
| 7 | Computer vision | CV | 27,228 | 23.3 |
| 8 | Deep learning | DL | 37,925 | 32.5 |

| 9 | Machine learning | ML | 76,106 | 65.2 |

(2) Classifying AI authors in AI preprint dataset

Based on the topics of preprints, we assigned an AI author to an AI subfield if the author's share of preprints in the subfield was higher than that of the AI authors' average. The statistical method we used guarantees that each author can be assigned to at least one AI subfield and takes the differences in subfield size into consideration (Sinatra et al., 2015). To remove the authors whose contributions to AI were marginal, we limited our analysis to 18,897 authors with more than 5 AI preprints. The detailed information on AI authors in the nine AI subfields is shown in Table 2. We observed that the largest two AI subfields were machine learning, with 9,385 scientists (49.6%), and deep learning, with 8,834 scientists (46.75%).

**Table 2**
The distribution of AI authors in the nine AI subfields.

| No. | # of subfields AI author working in | # of AI authors | Percentage (%) |
|---|---|---|---|
| 1 | NLP | 5742 | 30.39 |
| 2 | KR | 4231 | 22.39 |
| 3 | PS | 3885 | 20.56 |
| 4 | IR | 2584 | 13.67 |
| 5 | RO | 1344 | 7.11 |
| 6 | IA | 2352 | 12.45 |
| 7 | CV | 7838 | 41.48 |
| 8 | DL | 8834 | 46.75 |
| 9 | ML | 9385 | 49.66 |

One AI author can work in multiple AI subfields. To understand the pace of AI from the AI players more explicitly, we further counted the number of AI authors working in 1, 2, 3, 4, and, 5+ subfields (Table 3). We found that more than 75% AI authors worked in two (43.89%) or three (32.09%) subfields, 12.55% AI authors specialized in one subfield, and only 2.01% AI authors worked in five or more subfields.

**Table 3**
The distribution of AI authors in five categories.

| # of subfields AI author working in | # of AI authors | Percentage (%) |
|---|---|---|
| 1 | 2371 | 12.55 |
| 2 | 8294 | 43.89 |
| 3 | 6064 | 32.09 |
| 4 | 1789 | 9.47 |
| >=5 | 380 | 2.01 |

Besides, similar with the AI preprints, we also divided the AI authors into high-influential and low-influential AI authors according to their influence, which was represented by the value of "InfluentialCitationCount" from Semantic scholar. In particular, all the AI authors were ranked in a descending order by author influence, and the top 20% of them were marked as high-influential, last 40% as low-influential (Gayen, Bhavsar & Chandra, 2017).

*3.3. Indicators for characterizing the pace of AI innovations*

In this paper, we propose three bibliometric indicators to quantify the pace of AI innovations based on the bibliographic information of AI preprints in arXiv: Average Time Interval, Innovation Speed, and Update Speed. The details on these three indicators are as follows.

**Average Time Interval (ATI)**: Time Interval ($TI$) is defined as the time difference between two adjacent preprints. For a given year, we average the $TI$s for all two adjacent preprints as the $ATI$ of this year. Fig. 2 shows an example of 33,396 preprints with their submission times in arXiv during 2019. To compute the $ATI$ of the year 2019, we first chronologically sorted all these preprints by time and then calculated the TIs for all two adjacent preprints. Finally, we computed the average value of all the 33,395 $TI$s as the $ATI$ of the year 2019, that is, 0.262 hour, which means that in every 0.262 hours there was an AI preprint submitted to arXiv in 2019.

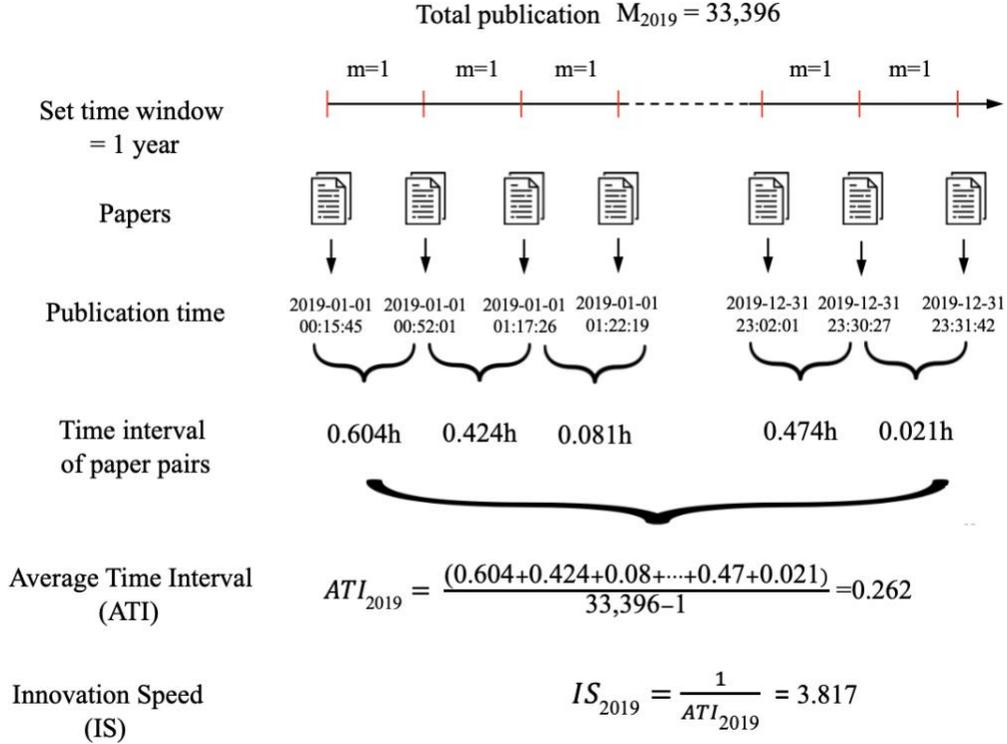

**Fig. 2.** An illustration of Innovation Speed.

Therefore, the mathematical definition of the ATI for a given year $Y$ is as follows.

$$ATI_Y = \frac{\sum_{j=2}^{M}(X_j - X_{j-1})}{M-1} \quad \text{hour(s)} \quad (1)$$

$M = \{x_1, x_2, x_3, \ldots, x_M\}$ means a set of all preprints submitted to arXiv in the year $Y$, in which preprints are ascendingly sorted by time. The $ATI$ can reflect the average duration of the time interval among preprints for a year. Furthermore, for new AI authors entering the AI field in the year $Y$, we define the $TI$ as the interval between the submission time of their first preprints in $Y$ for two adjacent new AI authors, to calculate the $ATI$ from the AI player perspective. An example for calculating the ATI of new AI authors for a given year is shown in Appendix B.

**Innovation Speed (IS)** is defined as the quantity of AI innovations within a unit time (usually an hour). The innovation speed of a specific year $Y$, $IS_Y$, is expressed as:

$$IS_Y = \frac{1}{ATI_Y} \quad \text{preprint(s)/hour} \quad (2)$$

For the year 2019, the IS is 3.817 preprint/hour (Fig. 2), indicating that 3.817 AI preprints were submitted to arXiv in every hour in 2019.

**Update Speed**: for each preprint, we define $T_{last}$ as the update time of its last version, $T_{initial}$ as its initial release time of the first version, and $N_v$ as the number of versions. Then, the $Update\ Speed$ ($US$) for an individual preprint is $\frac{N_V}{(T_{last}-T_{initial})}$. Therefore, in a given year Y, the update speed, $US_Y$, is denoted as follows.

$$US_Y = \frac{M-1}{\sum_{j=2}^{M}(\frac{(T_{last,j}-T_{initial,j})}{N_{v,j}}+\frac{(T_{last,(j-1)}-T_{initial,(j-1)})}{N_{v,j-1}})} \text{ update(s)/hour} \qquad (3)$$

*3.4 Analyzing the pace of AI innovations from three perspectives*

In this paper, we use the three proposed bibliometric indicators to depict the increasing velocity of innovations in AI from three perspectives, i.e., AI publications, AI players and AI updates. Firstly, AI publications can be considered as one of the most important forms of AI innovations according to the classical definition of innovation by Rogers (Rogers, 2010), indicating that we can use the production velocity of AI publications to represent the velocity of AI innovations. However, because of the long length of time lags from the submission to official publication, there is significant bias in using publications in journals and conferences to quantify the velocity of innovations. Thanks to the open science (Allen, Chris & David, 2019), AI publications in arXiv (i.e., AI preprints) with rapid dissemination mechanisms and fine-grained metadata, make it possible for us to quantify the pace of AI innovations. On the one hand, although the AI preprints in arXiv are not peer-reviewed, they have been reviewed by the domain moderators to guarantee their quality to some extent. A number of preprints have been eventually published in high-quality journals and conferences (Lin et al., 2020); some of the preprints have also been proved to be scientific breakthroughs, such as the proof of the Poincaré conjecture (Morgan, 2009). On the other hand, the submission and update history of the preprints in arXiv has been recorded at second level and the time lags from submission to open access are usually short and can be ignored. Consequently, the number of AI preprints over time and the Innovation Speed (IS) of different categories of AI preprints over time, are used to represent the pace of AI innovations from the perspective of AI publications.

Secondly, innovations depend heavily on talents (Chen & Huang, 2009). Gupta et al. (2006) argued that newcomers can significantly accelerate the innovation speed of a team by breeding novel ideas and diversifying research perspectives. The emerging growth of new authors in a scientific domain has also been proved to be able to improve the innovation ability of the domain and boost in innovation speed (Jones, 1995; Wang & Hicks, 2015). Therefore, the Innovation Speed (IS) of different categories of new AI authors (Appendix B) and the number of new AI authors over years are used to characterize the pace of AI innovations from the perspective of AI players.

Thirdly, trial-and-error is regarded as one of the most important elements of innovations (Azoulay, Graff, and Manso, 2011). Successful innovations usually have shorter and more frequent trial-and-error lifecycles. Donoso (2017) considered that it was unacceptable to omit trial and error when measuring innovation speed. Therefore, by defining trial and error in AI innovations as the process of refining, improving and updating the current version of AI preprints (Larivière et al., 2014), we use the Update Speed (US) of AI preprints to represent the pace of AI innovations from the perspective of AI updates. We also investigate the Average Time Interval (ATI) of AI preprints receiving their first citations over years, and the citation performance of AI preprints with different updated versions.

# 4. Results

## 4.1. The pulse of AI innovations

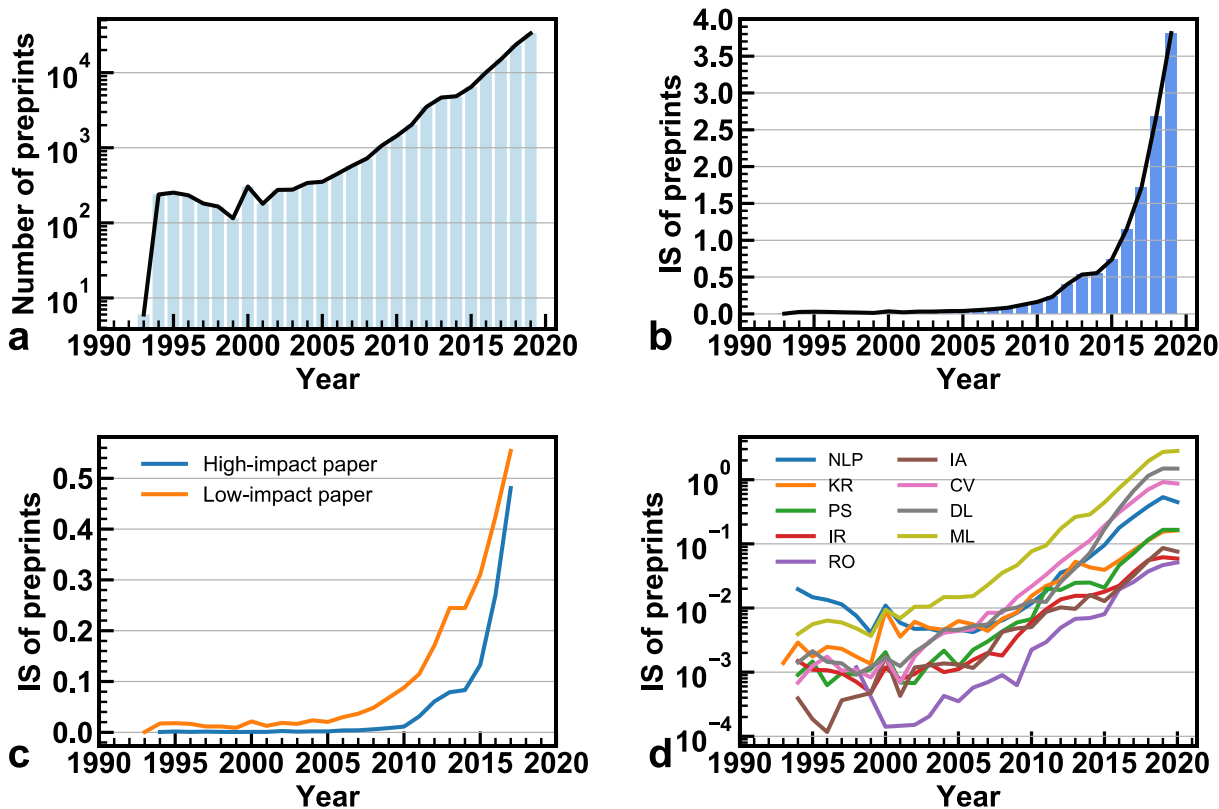

**Fig. 3.** The rapid AI innovation from the perspective of AI publications: (a) the annual number of AI preprints in arXiv (1993–2019); (b) the changes in the Innovation Speed of AI preprints in arXiv over time; (c) the changes in the Innovation Speed of high-impact (top 20%, blue line) and low-impact (last 40%, orange line) AI preprints; (d) the changes in the Innovation Speed of AI preprints in different AI subfields. (Note, NLP [Natural language processing] KR [Knowledge representation and reasoning], PS [Planning and scheduling], IR [Information retrieval], RO [Robotics], IA [Intelligent agents], CV [Computer Vision], DL [Deep Learning], and ML [Machine learning]).

Fig. 3a shows the annual preprints in arXiv between 1993 and 2019, which exhibit a clear tendency to increase over time. The number of AI preprints has roughly increased by an order of magnitude every 6 years; currently, more than 1,0000 AI preprints are submitted to arXiv each year, compared with the fewer than 30 annual submissions roughly 25 years ago, making for a rapid growth of AI preprints. Fig. 3b indicates that the Innovation Speed of AI preprints accelerates swiftly. Specifically, up to 1994, it took more than 39 hours to see an AI preprint submitted to arXiv, whereas in 2019, there were 3.817 AI preprints submitted to arXiv every hour.

Similarly rapid growth also occurred in high-impact (top 20%) and low-impact (last 40%) AI preprints (Fig. 3c), both of whose innovation speeds have exhibited a sharp rise since 2007, indicating that it took much less time to see a high-impact or low-impact AI preprint submitted to arXiv. For instance, the Average Time Interval of high-impact AI preprints in 1994 was about 67 days (IS ≈ 0.000624 preprints/hour), while it has dramatically decreased to less than 10 hours (IS > 0.1 preprints/hour) since 2015. Furthermore, when comparing the Innovation Speed of these two curves, the submission of a high-impact AI preprint to arXiv requires more time than a low-impact one. However, the gap in the Innovation Speed between high-impact and low-impact AI preprints has been gradually narrowed.

To obtain a detailed understanding of the proliferation of AI pace from the perspective of AI publications, we investigated the changes in the Innovation Speed of AI preprints in different AI subfields from 1993–2019 (Fig. 3d), demonstrating that it takes less time to see an AI preprint submitted to arXiv in all subfields. Specifically, the Average Time Interval for preprints in machine learning has sharply decreased from 11 days (IS = 0.0039 preprints/hour) in 1994 to only 0.513 hours (IS = 2.712 preprints/hour) in 2019. Similarly, it took around 30 days (IS = 0.0014 preprint/hour) to see a new deep learning–related preprint submitted to arXiv in 1994, but only 0.87 hours (IS = 1.49 preprints/hour) in 2019.

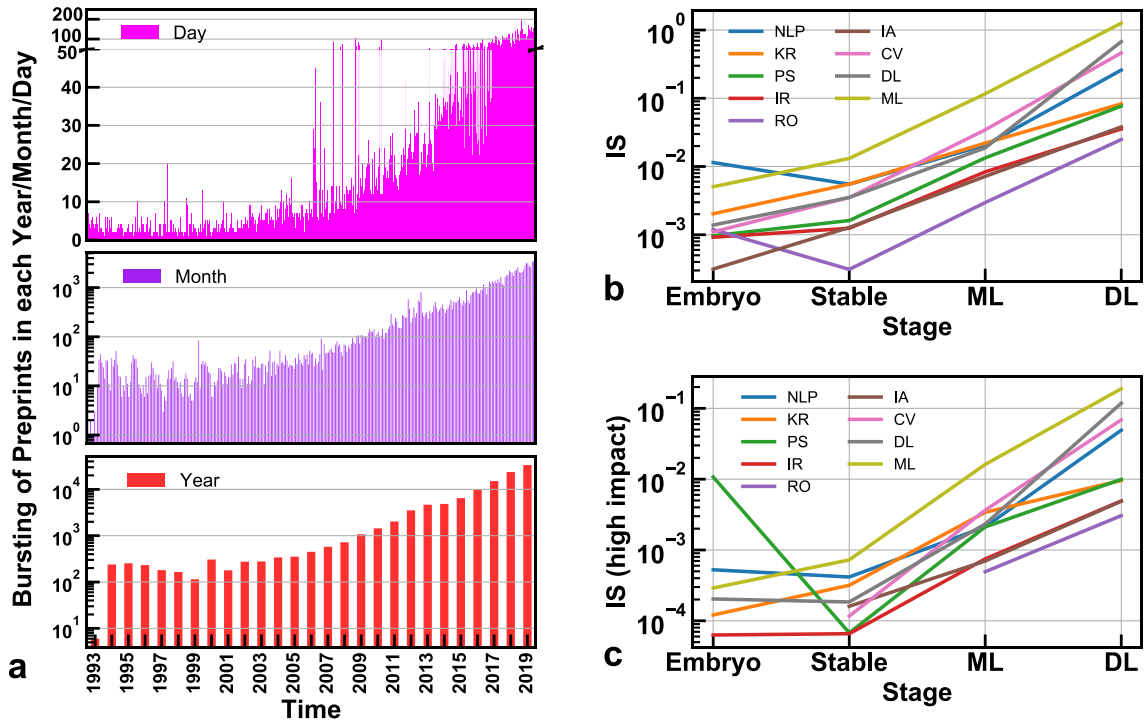

**Fig. 4.** The innovation speed of AI from the perspective of AI preprints: (a) the bursting of AI preprints for each year/month/day; (b) the changes in the Innovation Speed of AI preprints in different AI subfields across all four stages of AI; (c) the changes in the Innovation Speed of high-impact preprints in different AI subfields across all four stages of AI.

**Table 4**

The detailed information for the AI preprints between 1993 and 2019. Note: "Days Coverage" refers to the percentage of days in a year that have at least one AI preprint submitted to arXiv.

| Stage | Time range | # of Preprints (yearly) | # of Preprints (monthly) | Days Coverage |
|---|---|---|---|---|
| Embryo | 1993–1999 | $\approx 10^2$ | $\approx 10^1$ | 27% |
| Stable | 2000–2007 | $\approx 10^2 - 10^3$ | $\approx 10^1$ | 48% |
| Machine learning (ML) | 2008–2013 | $\approx 10^3 - 10^4$ | $\approx 10^2$ | 92% |
| Deep learning (DL) | 2014–2019 | $\approx 10^4$ | $\approx 10^3$ | 100% |

The number of AI preprints across all the four stages is shown in Fig. 4a and Table 4. The yearly, monthly, and daily number of AI preprints also increase over time. For instance, in the embryo stage (1993–1999), the yearly and monthly number of AI preprints were about 100 and 10, while in the deep learning stage (2014–2019), they have grown into around 10,000 and 1,000, respectively. Furthermore, from 1993–1999, there were 98 days (27% of a year) in each of which at least one AI preprint was submitted to arXiv; however, between 2014 and 2019, new AI preprints were submitted every day of each year.

Fig. 4b represents the changes in the Innovation Speed of AI preprints in different subfields over four stages. Innovation speeds for all subfields exhibit an increasing trend overall, although the maximum speeds for different AI subfields were different. Specifically, the Innovation Speed of machine learning (ML), deep learning (DL) and computer vision (CV) ranked in the top three places of the fourth stages (2014–2019). The Innovation Speed of Nature Language Processing (NLP) had ranked first at the embryo stage, but it was surpassed by that of CV and DL in the machine-learning and deep-learning stage, respectively. Moreover, the Innovation Speed of DL showed a spike and ranked second in the last stage. The Innovation speeds of NLP and Robotics show an evident decline in the first stage, which may indicate the failure of traditional AI and the advent of an AI winter (Hendler, 2008).

Fig. 4c shows the changes in Innovation Speed among different AI subfields for high-impact AI preprints. Overall, the Innovation Speed of all AI subfields exhibits an increasing trend in four stages, except for Planning and Scheduling (PS), showing a decreasing–increasing trend. This illustrates the rise and fall in the historical evolutionary path of PS research during that period. The growth trend starts to slow down in the last stage, except for DL, exhibiting a swift increase during the machine-learning stage and establishing its dominate position at the deep-learning stage (Arel, Rose, & Karnowski, 2010).

*4.2. Rapid growth of AI players*

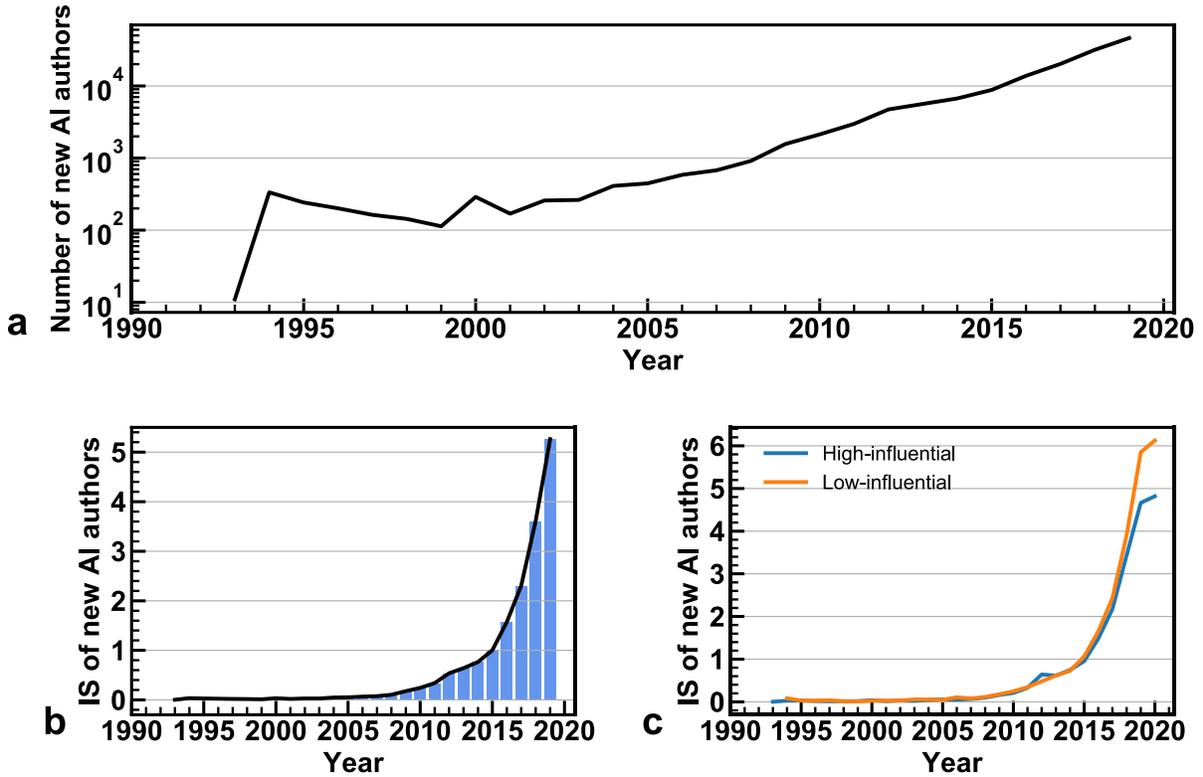

**Fig. 5.** The rapid innovation of AI from the perspective of AI players: (a) the changes in the annual number of new AI authors in AI preprint dataset over time; (b) the changes in the Innovation Speed of new AI authors in AI preprint dataset over time; (c) the changes in the Innovation Speed for high-influential (top 20%) and low-influential (last 40%) new AI authors over time.

Nonstop waves of AI authors continue spurring the AI innovations. As shown in Fig. 5a, there were 46,097 new authors entering into the field of AI according to the AI preprint dataset in 2019. The growth speed of the annual number of new AI authors in the year 2019 is much faster than that in 1995. Fig. 5b shows the changes in the Innovation Speed for new authors entering into the AI field over the years and months based on the AI preprint dataset. The Innovation Speed for a new author entering into the AI field has increased from less than 0.03 in 1994 to around 5.26 in 2019, indicating that there were about 5.26 new authors entering into the field of AI every hour in 2019, which is about 175 times that of the speed in the 1990s. Meanwhile, the Innovation Speed of new AI authors between the high-influential (top 20%) and low-influential (last 40%) new AI authors are roughly same, and both exhibit a sharp increasing trend, especially since 2010 (Fig. 5c).

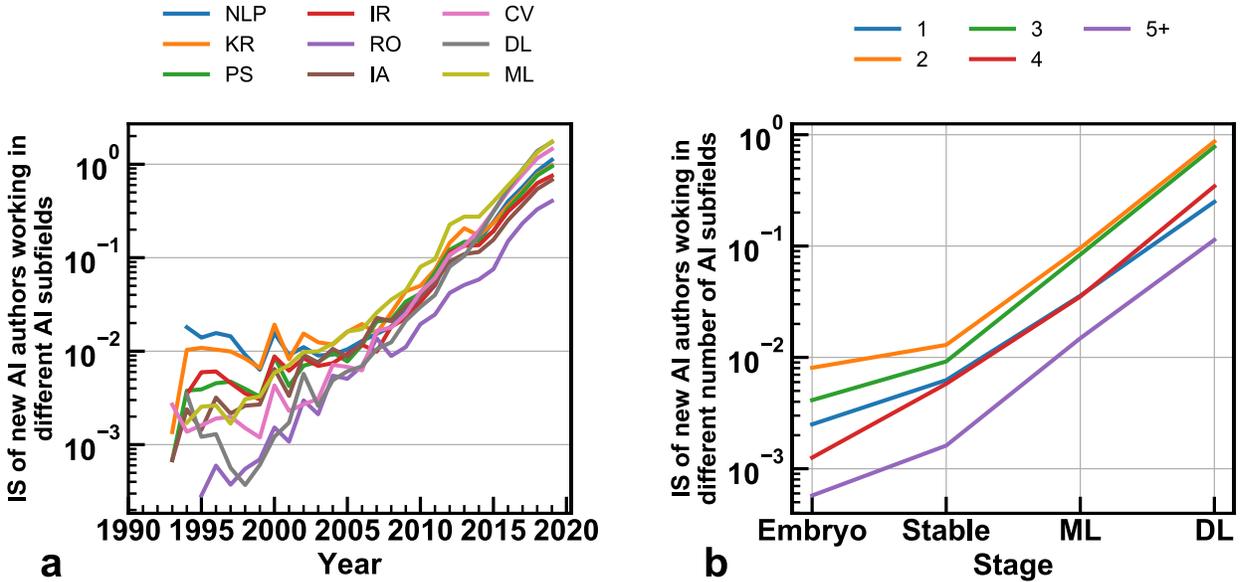

**Fig. 6.** The innovation speed of AI from the perspective of AI players in arXiv**:** (a) the changes in the Innovation Speed of new AI authors in different AI subfields over time; (b) the changes in the Innovation Speed of new AI authors working in different number of AI subfields over stages.

To achieve a comprehensive understanding of the AI innovation speed from the perspective of AI players, we calculated the changes in the Innovation Speed of new AI authors for nine different AI subfields (Fig. 6a). The Innovation speeds of new AI authors in the different AI subfields varies and are small in the embryo stage (1993–1999); however, gaps among different subfields gradually disappear as time passes, and all curves exhibit a clear increasing trend over time. Like with AI preprints, the Innovation Speed of new AI authors in the subfields of machine learning, deep learning, and computer vision ranked first, second and third, respectively.

We analyzed the Innovation speeds of new AI authors working in different number (that is, 1, 2, 3, 4, 5+) of AI fields over stages (Fig. 6b). All five curves exhibit clear increases over the different AI stages. New AI authors working in 2 or 3 AI subfields had the first and second Innovation Speed over four stages, while the Innovation Speed of authors who specialized in one AI subfield ranked third at the embryo stage, then was surpassed by new authors working in 4 AI subfields at the stable stage and ended at the fourth in the deep-learning stage. In addition, the Innovation Speed of new authors working in 5 or more AI subfields has been consistently been the slowest, despite having exhibited an evident growth over stages.

## 4.3. Trial and error as innovation improvement

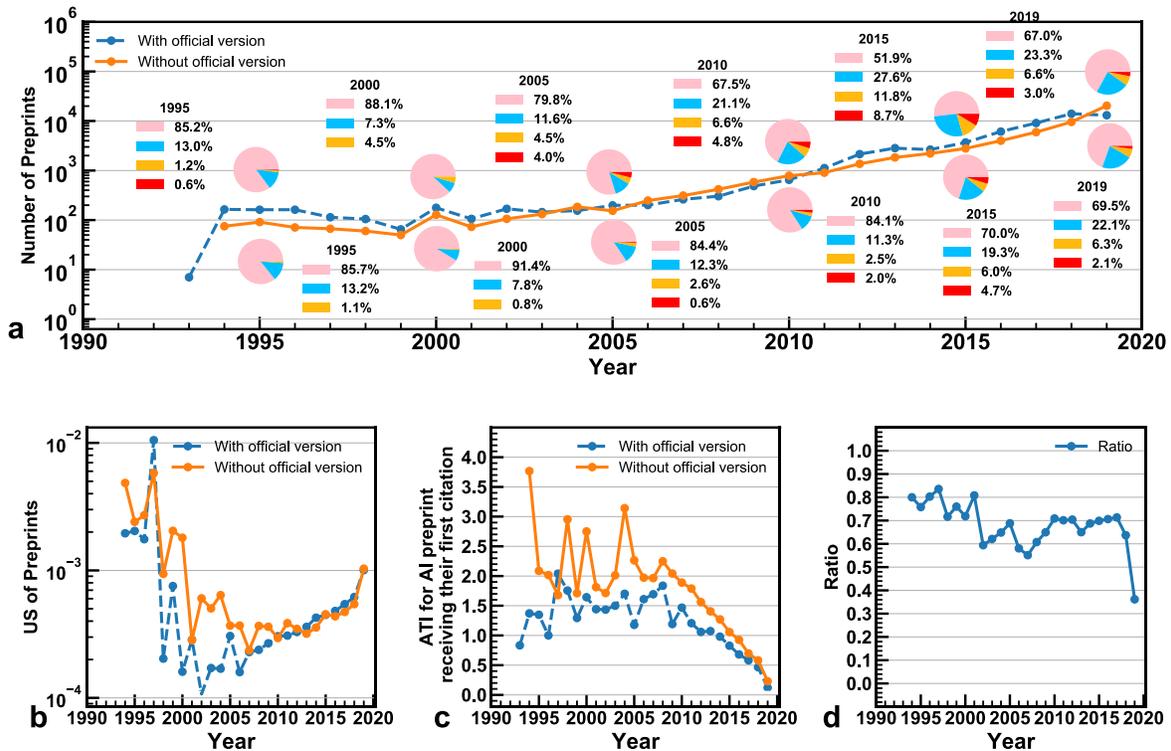

**Fig. 7**. The rapid AI innovation from the perspective of trial and error: (a) the changes in the annual number of AI preprints with and without official version, respectively (note that the pie charts represent the percentages of preprints with a different number of versions, including [1] 1 version [pink]; [2] 2 versions [blue]; [3] 3 versions [yellow]; and [4] 4 or more [red]); (b) the changes in the Update Speed of AI preprints with and without official version, respectively; (c) the changes in the Average Time Interval for AI preprints receiving their first citation over years; and (d) the changes in the ratio of AI preprints having one or more citations over years.

The urge to be the first or to gain public acknowledgement through early publications prompts authors to continue updating their papers. As shown in Fig. 7a, the annual number of AI preprints with or without official versions have both been increasing overall. The number of updated versions of AI preprints exhibits a clear increasing trend, where 33% AI preprints with official versions and 33.5% AI preprints without official versions have been updated at least twice in 2019, while few AI preprints had updated versions in the 1990s. In addition, no differences were observed between the changes in the number of updated versions of AI preprints with and without official versions.

Fig. 7b shows the changes in Update Speed for AI preprints with at least one update, in which the blue dotted line represents preprints with official versions and the orange line represents preprints without official versions. The Update Speed of both groups exhibits an uneven but clear upward trend between 2000 and 2019.

In 2019, it cost less than 41 days (995 hours, US = 0.001005) on average to see an updated version of AI preprints submitted to arXiv, and there were no differences in the Update Speed between AI preprints with and without official versions.

To understand the pace of AI from the perspective of trial and error, it is critical to investigate the citation impact of AI preprints, which is an important motivation of trial and error. According to the citation counts within 3 years, around 70% of AI preprints in arXiv received one or more citations each year between 1993 and 2017 (Fig. 7d). Fig. 7c reveals a clear decreasing trend over time with respect to the Average Time Interval for an AI preprint in both groups to receive its first citation. Specifically, the Average Time Interval to receive the first citation for most of the preprints in both groups is around 1.5 years during the stable stage (2000–2007), which then declined to about 1 year during the machine-learning stage (2008–2013) and finally dropped to 0.22 year for AI preprints with official versions and 0.12 year for AI preprints without official versions in 2019. In addition, we can find that the Average Time Interval to receive the first citations for AI preprints with official versions has been longer than that of AI preprints without official versions.

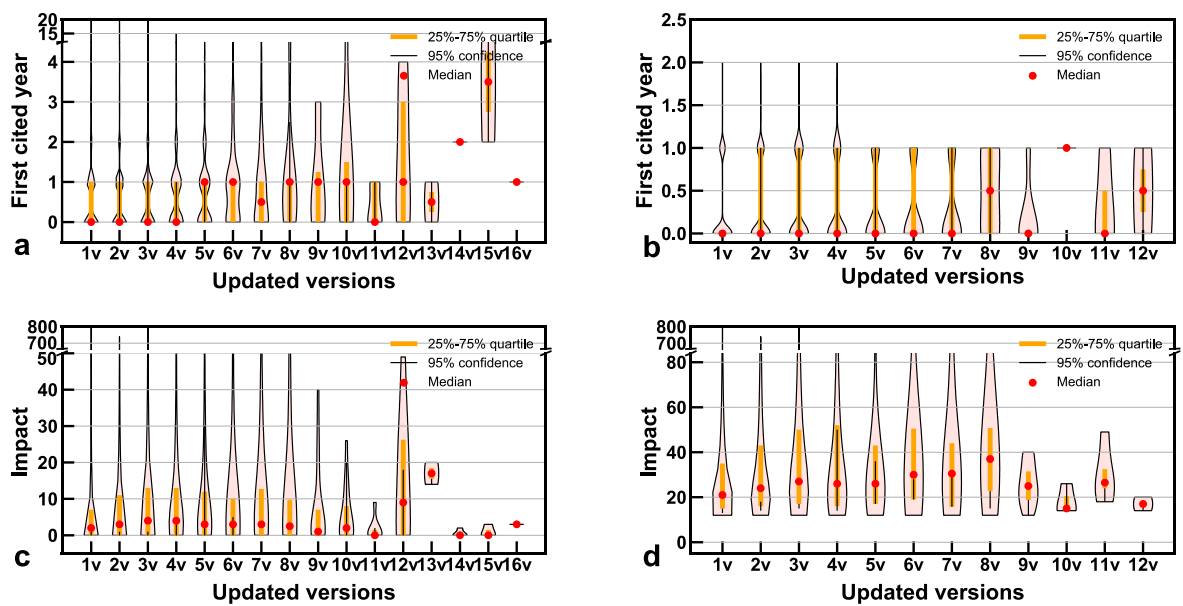

**Fig. 8.** Short-term and long-term performance of AI preprints with different updated versions: (a) the first cited year for AI preprints with a different number of updated versions before its official publication; (b) the first cited year for high-impact (top 20%) AI preprints with a different number of updated versions before its official publication; (c) the citation counts for low-impact (last 40%) AI preprints with a different number of updated versions before its official publication; and (d) the citation counts for AI preprints with a different number of updated versions before its official publications.

Next, for AI preprints with official versions in arXiv, we looked into the relationship between their number of updated versions and the time to receive their first citations. Note that if an AI preprint has N updated versions, then it has (N+1) versions submitted to arXiv. Fig. 8a shows that all AI preprints with fewer than 5 updated versions can averagely receive their first citation at the fastest speed, i.e., in the same year of their submission; however, for the high-impact (top 20%) ones (Fig. 8b), AI preprints with more than 5 updated versions, such as 6, 7, 9, and 11 updated versions, can also receive their first citation in the same year. Overall, it is faster for high-impact AI preprints to receive their first citations, and AI preprints with fewer updated versions tend to receive their first citations earlier.

The aforementioned time to receive their first citations is the short-term performance of AI preprints, we here also investigate the long-term performance (i.e., citation counts) of AI preprints with different number of updated versions (Aksnes et al., 2019). As shown in Fig. 8c, most AI preprints with fewer than 12 updated versions have received around 5 citations, and most high-impact AI preprints have received over 20 citations (Fig. 8d). There exists a slight advantage in gaining more citations for high-impact AI preprints with more updated versions if to have fewer than 9 updated versions. However, when high-impact AI preprints have more than 8 versions, their citation counts decline.

## 5. Discussion and conclusions

This article explores the pace of AI from the perspective of AI papers, AI players, and trial and error, using three measures: Average Time Interval, Innovation Speed, and Update Speed. On the one hand, time-based (i.e., day/month/year) and stage-based (i.e., embryo, stable, machine learning, and deep learning stages) longitudinal analyses are employed to track the innovation speed of AI. On the other hand, the horizontal analysis based on impact (i.e., high-impact and low-impact AI papers), influence (high-influential and low-influential AI authors) and different AI subfields, provide the nuanced details about the pace of AI.

Also, this article includes two reliable data sources of AI research to shed light on the pace of AI innovation from three different perspectives. The findings reveal that more and more talents are entering into the field of AI, which may directly stimulate the drastic growth in the number of AI innovations (e.g., AI papers). Meanwhile, the findings suggest that the authors might not benefit from the frequent and quick attempts to update the versions of AI preprints. Hence, AI authors should be encouraged to pay more attention to the quality of their papers but not their speed.

Methodologically, this study first provided a framework for investigating the innovation pace of a fast-moving research field by analyzing the input, output and improvement of innovations based on the fine-grained metadata recorded in preprints and publications (i.e., arXiv and Semantic Scholar). This framework could be adopted and generalized to measure the pace of innovations in other fast-moving research fields, such as cancer research and nano science. Second, in this study, we developed three bibliometric indicators, i.e., Average Time Interval (ATI), Innovation Speed (IS) and Updated Speed (US); and we showcased how to use them to measure the pace of innovations in the field of AI from three perspectives. The results demonstrated that ATI, IS and US can effectively reflect the time intervals of newly emerged AI innovations, the fast growth of AI publications and working force, and the rate of trial and error in AI innovations, respectively. On the one hand, these indicators can be applied to quantify the pace of innovations in other domains. On the other hand, they can also be adopted and modified for other research tasks, such as, measuring the time intervals of emerging research topics, and depicting the temporal trends of scientific concepts.

Third, we used the updates of preprints to represent the trial and error of the innovation process in a research field, which provides a novel way to quantify the trial and error of scientific innovations from the perspective of bibliometrics. Fourth, using the arXiv dataset to understand the pace of innovation can be also a methodological contribution of this study to the related fields. Compared to other bibliographic datasets, preprints in arXiv has short time lags from submission to be posted and these actions are all recorded in second level, which make the arXiv data a better choice for measuring the pace of innovation in a research field.

Nevertheless, the current study has several limitations. First, it only considers the authors of AI papers, which ignores the talents in AI companies (Chaturvedula et al, 2019; Seeber et al., 2020). Second, the findings on the innovation speed of AI are mainly based on the coverage and correctness of the arXiv data, as it records the submission time of AI preprints at seconds. Although the arXiv data majorly cover publications on computer science (Kim, 2019; Larivière et al., 2014), data about other fields should be considered (e.g., bio-arXiv). Specifically, the arXiv has different coverage from other academic databases, such as Scopus, DBLP, and Web of Science. Some AI conferences (e.g., ACM SIGKDD 2019) suggested that articles should not be published on the arXiv before the review decision because of the double-blind rule. These factors might affect the results of this paper. Therefore, the findings of this paper should be interpreted to represent only the current dataset as it is. Third, although this paper shows the pace of AI innovations from different perspectives, it does not analyze factors affecting this speed. In our future work, we will further explore the AI innovation speed from the perspective of innovation diffusion, team collaboration, labor division, and diversity. Eventually, studying the relationships between the speed of innovation and various factors, such as labor division, team diversity, and diffusion models, will help us to better understand AI and the pace of its development, and, ideally, facilitate policymakers' decision-making.

## Acknowledgement

This work was supported by the National Social Science Foundation of China (71420107026,91646206) and the National Research Foundation of Korea (No. NRF-2019R1A2C2002577). This work was also supported by the Yonsei University Research Grant of 2020. The support provided by the China Scholarship Council (CSC) during a visit by Xin Li to the University of Texas at Austin is acknowledged (No. 201806270047). We thank Xiaoran Yan for his great help in the process of data collection. We also thank the Indiana University's Supercomputer service of Cabornate and Karst, which provided abundant computing resources for this work.

# Appendix A. Topic words for classifying AI preprints to nine subfields

**Table A.1.** Topic words for classifying AI preprints to nine subfields

| No. | Paper Categories | Topic words |
|---|---|---|
| 1 | Natural language processing | "Natural language processing," "Information extraction," "Machine translation," "Discourse, dialogue and pragmatics," "Natural language generation," "Speech recognition," "Lexical semantics," "Phonology/morphology," "Tokenization," "Stemming," "Lemmatization," "Stop words," "Parts-of-speech tagging," "POS tagging," "Statistical language modeling," "Bag of words," "n-grams," "Regular expressions," "Sentiment analysis," "Ontology induction," "Question answering," "Semantics and summarization," "Speech processing," "Text classification," etc. |
| 2 | Knowledge representation and reasoning | "Knowledge representation and reasoning," "Description logics," "Semantic networks," "Nonmonotonic, default reasoning and belief revision," "Probabilistic reasoning," "Vagueness and fuzzy logic," "Causal reasoning and diagnostics," "Temporal reasoning," "Logic programming and answer set programming," "Spatial and physical reasoning," "Reasoning about belief and knowledge," "Automated reasoning and theorem proving," "Case-based reasoning," "Common-sense reasoning," "Belief change," "Computational complexity of reasoning," "Description logic(s)," "First-order predicate," "First-order logic" "Diagnosis and abductive reasoning," "Geometric reasoning," "Spatial reasoning," "Temporal reasoning," "Knowledge representation languages," "Logic programming," "Nonmonotonic reasoning,", "Qualitative reasoning," "Reasoning with/about beliefs," "Knowledge representation" , "Word2Vec," "Doc2Vec," "Structure2vec," etc. |
| 3 | Planning and scheduling | "Planning and scheduling," "Planning," "Scheduling," "Planning for deterministic actions," "Planning under uncertainty," "Multi-agent planning," "Planning with abstraction and generalization," "Robotic planning," "Evolutionary robotics," "Activity and plan recognition," "Deterministic planning," "Learning models for planning and diagnosis," "Markov models of environments," "Mixed discrete planning," "Mixed continuous planning," "Model-based reasoning," "Plan execution and monitoring," "Probabilistic planning," "Replanning and plan repair," "Temporal planning," etc. |
| 4 | Information retrieval (Search methodologies) | "Heuristic function construction," "Heuristic function," "Heuristic Search," "Discrete space search," "Continuous space search," "Randomized search," "Random search," "Game tree search," "Distributed tree search," "Distributed Search," "State space search," "Monte carlo tree search," "Abstraction and micro-operators," "Search with partial observations," "Metareasoning and Metaheuristics," etc. |
| 5 | Robotics | "Robotic planning," "Evolutionary robotic(s)," "Computational control theory," "Motion control," "Motion path planning," "Motion planning," "Admission control," "Air traffic control," "Computational control theory," "Flow control," "Linear quadratic gaussian control," "Remote control," "Robot control," "Robotic planning," "Sliding mode control," etc. |
| 6 | Intelligent agents | "Intelligent agent(s)," "Mobile agent(s)," "Cooperation and coordination," "Multi-agent system(s)," "Ad-Hoc teamwork," "Agent/AI theories and architectures," "Agent-based simulation and emergent behavior," "Agent communication," "Coordination and collaboration," "Distributed problem solving," "Opponent modeling," "Mechanism design," "Multiagent learning," "Multiagent planning," "Evaluation and analysis (Multiagent Systems)," "Decentralized Artificial Intelligence," etc. |
| 7 | Computer vision | "Computer vision," "Face and gesture recognition," "Image and video retrieval," "Language and vision," "Motion estimation," "Motion capture," "Interest point and salient region detections," "Image segmentation," "Video segmentation," "Shape inference," "Object identification," "Statistical methods and learning," "Scene reconstruction," "Scene anomaly detection," "Scene understanding," "Video tracking," "Visual inspection," "Vision for robotics," "Active vision," "3D pose estimation," "3D imaging," "Image restoration," "Camera calibration," "Epipolar geometry," "Computational photography," "Hyperspectral imaging," "Image representations," "Shape representations," "Appearance and texture representations," etc. |
| 8 | Deep learning | "Deep learning," "Deep neural network(s)," "Convolution/Convolutional neural network(s)/CNN(s)/ConvNet," "Long short term memory/LSTM," "Recurrent neural network(s)/RNN(s)/," "Bidirectional recurrent neural network(s)," "Deep belief network(s)," "Restricted Boltzmann machine(s)/RBM(s)," "Boltzmann machine(s)," "Generative adversarial network(s) (GAN(s))," "Autoencoder," "Variational auto-encoder," "Recursive neural network(s)," "Bidirectional recurrent neural network(s)," "Graph |
| | | convolutional network(s)," "Regional convolutional neural network(s)," "Deep reinforcement learning," "Deep Q network(s)," "Dueling deep Q network(s)," "Double deep Q network(s)," "Word2Vec," "Doc2Vec," "Structure2vec," etc. |
| 9 | Machine learning | "Machine learning," "Active learning," "Bayesian learning," "Causal learning," "Classification," "regression trees," "Kernel methods," "Support vector machine(s)/SVM(s)," "Gaussian process(es)," "Neural network(s)," "Inductive logic learning," "Statistical relational learning," "Maximum likelihood modeling," "Maximum entropy modeling," "Maximum a posteriori modeling," "Mixture model(s)," "Latent variable model(s)," "Bayesian network model(s)," "Bayesian/Bayes network(s)," "Perceptron algorithm," "Non-negative matrix factorization," "Factor analysis," "Principal component analysis," "Canonical correlation analysis," "Latent dirichlet allocation," "Latent Semantic Indexing" ,"Rule learning," "Instance-based learning," "Markov decision process(es)," "Partially-observable markov decision process(es)," "Stochastic game(s)," "Genetic algorithm(s)," "Clustering," "Dimensionality reduction," "Feature selection," "Ensemble methods," "Evaluation and analysis (Machine Learning)," "Evolutionary learning," "Feature Construction/Reformulation," "Learning preferences or rankings," "Learning theory," "Multi-instance/Multi-label/Multi-view learning," "Probabilistic graphical model," Probabilistic directed acyclic graphical model," "Recommender system(s)," "Relational learning," " Semi-supervised learning," "Supervised learning," "Unsupervised learning," "Structured prediction," "Time-Series stream(s)," "Data stream(s)," "Transfer learning," "Adaptation learning," "Multi-task learning," ""K nearest neighbor/KNN/K-nearest neighbor/K-NN," "Belief network(s)," "Cellular nonlinear network(s)," "Ant colony optimization algorithm," "Particle swarm optimization," "Particle Swarm Optimization," "Artificial bee colony algorithm," "Latent dirichlet allocation/LDA," "Probabilistic latent semantic analysis/PLSA, etc. |

**Appendix B. An illustration for calculating the Innovation Speed of new AI authors.**

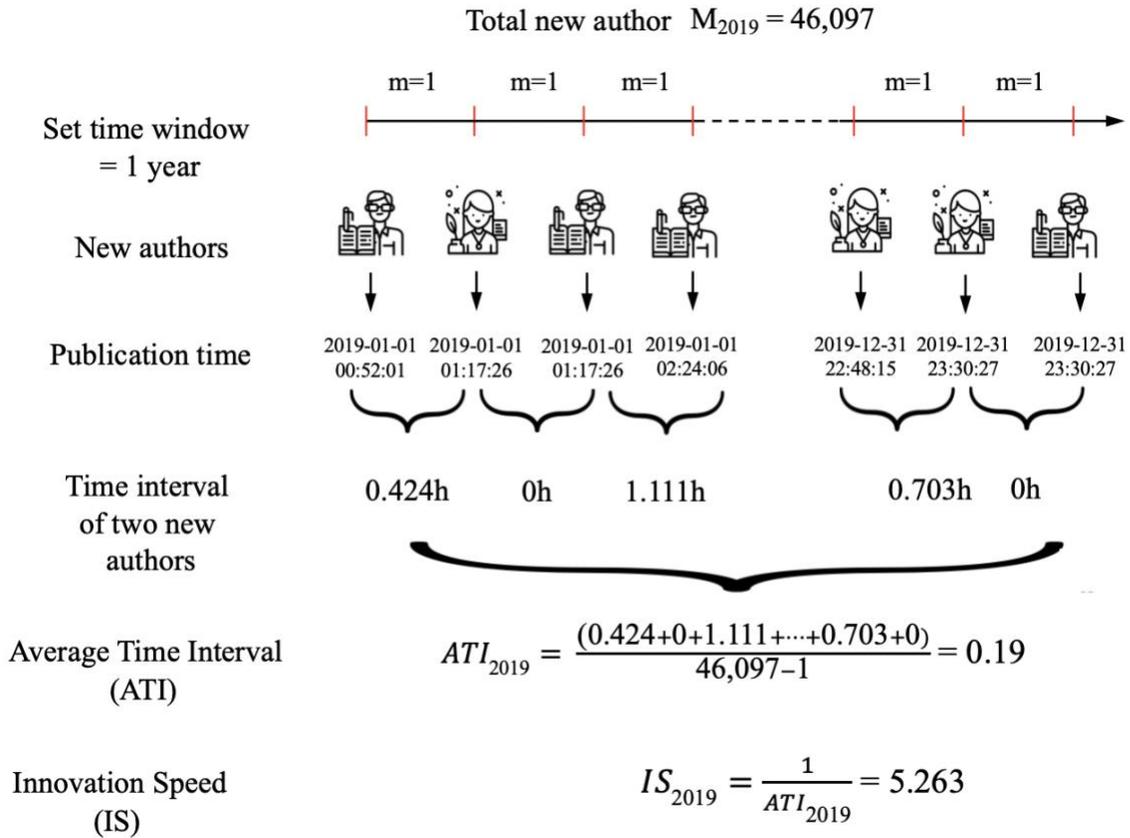

**Fig. B.1.** An illustration for calculating the Innovation Speed of new AI authors.

# Appendix C. The changes in the number of AI authors per preprint and the number of AI preprints per author over time.

The mean number of authors per AI preprint shows a clear increasing trend over time (Fig. C.1), however, the mean number of preprints per author has not evidently grown (Fig. C.2). This indicates that the growing number of AI authors and the collaborations between them could be the causes of the growing number of AI preprints.

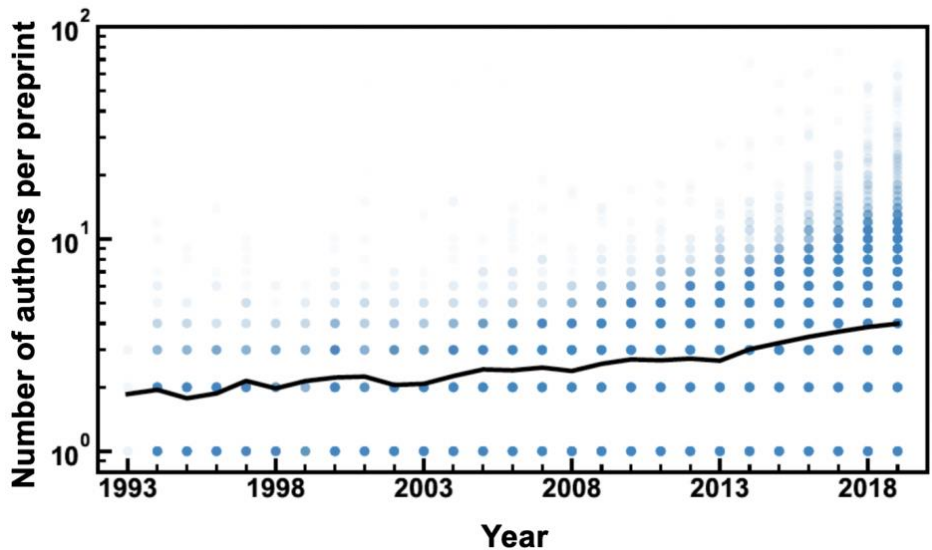

**Fig. C.1.** The changes in the number of authors per preprint over time. Note, the black line represents the changes in the mean number of authors per preprint over time. A blue circle represents a set of preprints sharing the same coordinates; the greater the number of preprints, the bluer the color.

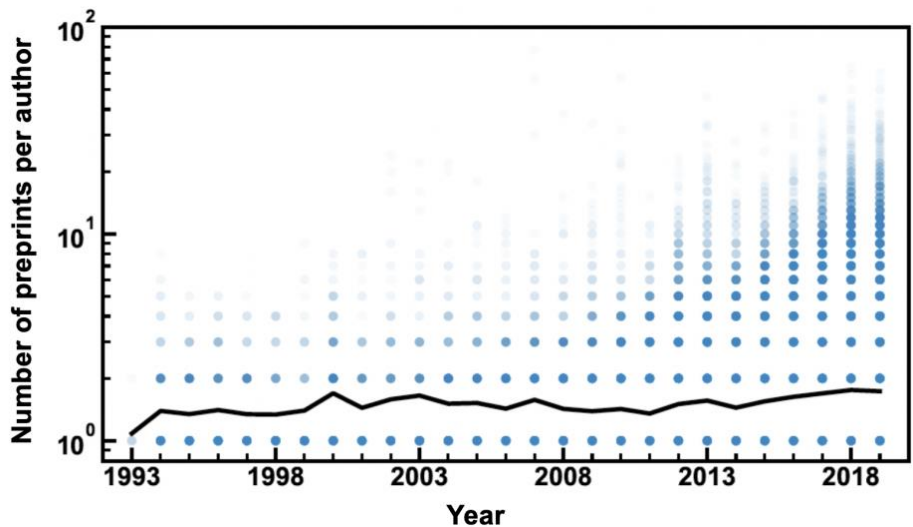

**Fig. C.2.** The changes in the number of preprints per author over time. Note, the black line represents the changes in the mean number of preprints per author over time. A blue circle represents a set of authors sharing the same coordinates; the greater the number of authors, the bluer the color.

There are total 157,856 AI authors who have submitted one or more AI preprints to arXiv from 1993 to 2019, in which 1,498 authors are "old" authors (who have submitted AI preprints to arXiv before 2000) and 156,358 authors are "new authors" (who haven't submitted any AI preprints to arXiv before 2000).

As shown in Fig. C.3, for old authors, the productivity (the number of preprints per unique author) has a clear growing trend with fluctuations over time. There is a significant peak occurred in 2013 for the productivity of old authors, which may be caused by the emerging of deep learning (Russakovsky et al., 2015).

Fig. C.4 shows the productivity of new authors. Before 2013, the mean productivity of these new authors has kept stable with small fluctuations; then, it showed a slight increase from 2013 to 2019. The productivity of new authors has been less than that of the old authors all the time (2001-2019). Hence, not only the growing number of new authors but also the growing productivity of old authors, has contributed to the swift increasing number of AI preprints.

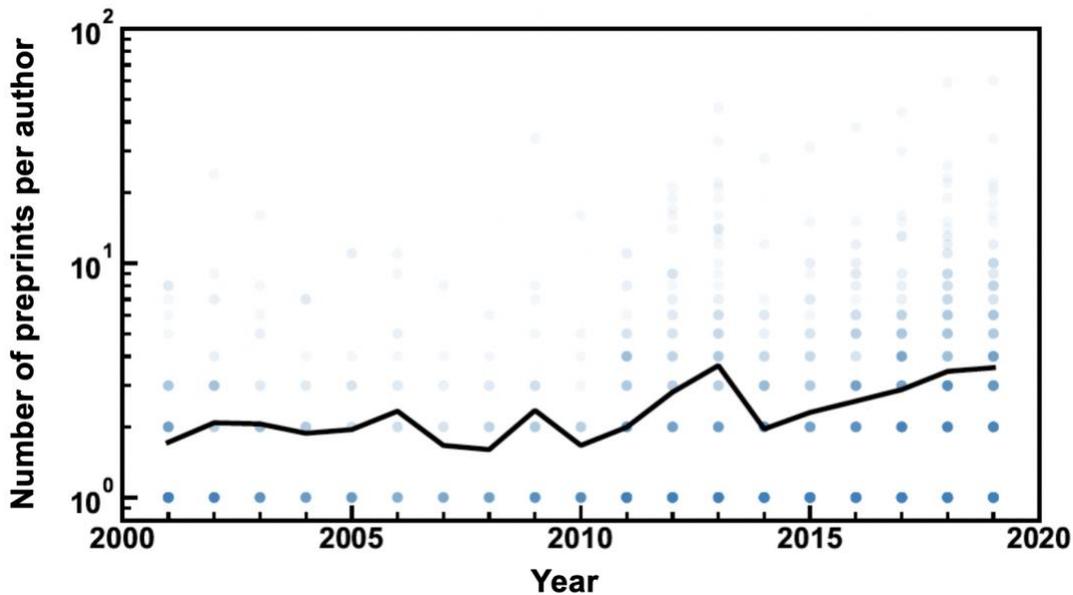

**Fig. C.3.** For "old" authors (who have submitted AI preprints to arXiv before 2000), the changes in the number of preprints per author over time. Note, the black line represents the changes in the mean number of preprints per author over time. A blue circle represents a set of authors sharing the same coordinates; the greater the number of authors, the bluer the color.

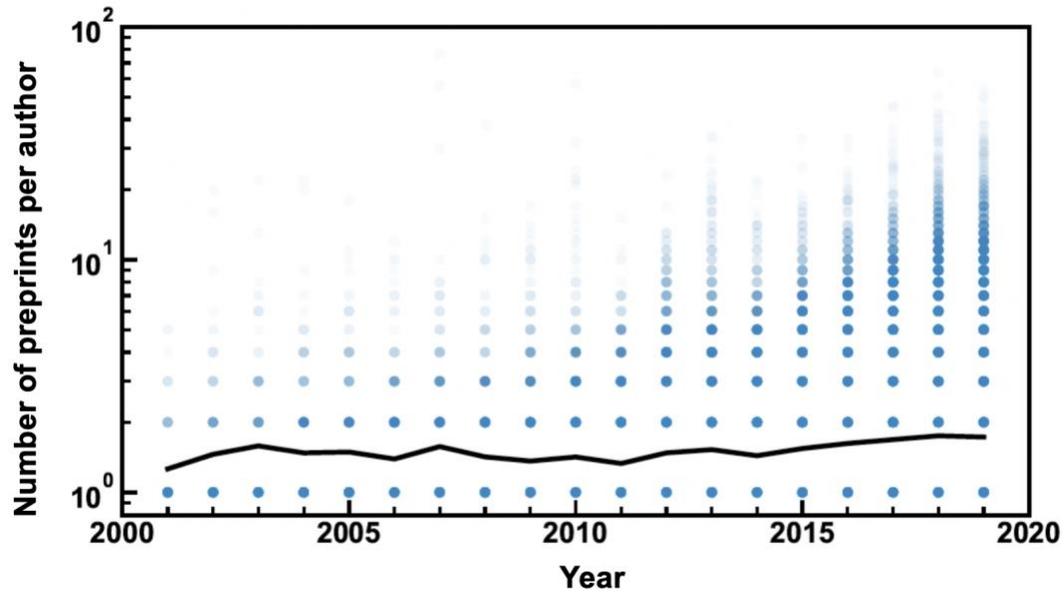

**Fig. C.4.** For "new" authors (who haven't submitted AI preprints to arXiv before 2000), the changes in the number of preprints per author over time. Note, the black line represents the changes in the mean number of preprints per author over time. A blue circle represents a set of authors sharing the same coordinates; the greater the number of authors, the bluer the color.

# Appendix D. The number of AI authors per preprint over different updated versions.

As shown in the Fig. D.1, when the number of updated versions is less than 7, there is no evident relationship between the number of authors per preprint and their updated versions. When the number of updated versions is more than 7 and less than 10, the number of authors per preprint showed a decreasing trend. When the number of updated versions is more than 10, the changes in the number of authors per preprints is irregular. This is due to the number of preprints that have been updated for more than 10 times is rather small. Hence, multiple authors should not be the reason of multiple updates in arXiv.

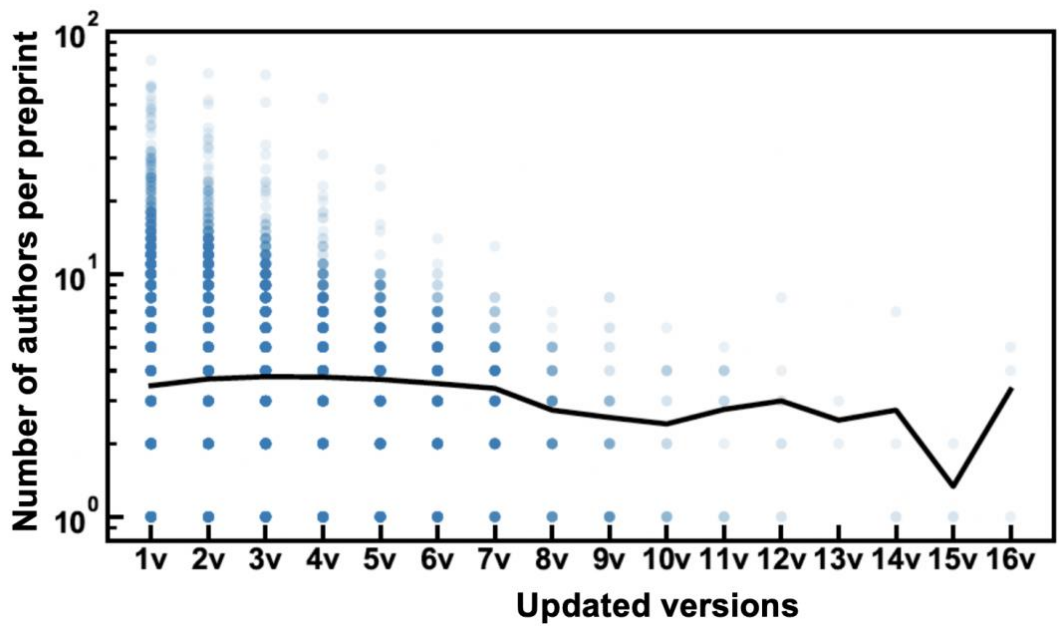

**Fig. D. 1.** The number of AI authors per preprint over different updated versions. A blue circle represents a set of preprints sharing the same coordinates; the greater the number of preprints, the bluer the color.